\documentclass{IEEEtran}

\usepackage[style=ieee, backend=bibtex, doi=false, url=false, isbn=false, date=year]{biblatex}
\AtBeginBibliography{\small}
\AtEveryBibitem{\clearlist{language}}
\AtEveryBibitem{ \ifentrytype{misc}{}{\clearfield{note}}}

\usepackage{amsmath,amssymb,amsfonts}
\usepackage{algorithmic}
\usepackage{graphicx}
\usepackage{textcomp}

\usepackage{nccmath}
\usepackage{mathrsfs}
\usepackage{mathtools}
\usepackage{amsthm}
\usepackage{graphicx}

\usepackage{textcomp}
\usepackage{subcaption}
\usepackage{url}

\usepackage[american]{circuitikz}

\usepackage{hyperref}
\hypersetup{
	colorlinks=true
}

\newcommand{\R}{\mathbb{R}}
\newcommand{\I}{\mathbf{I}}

\newcommand{\C}{\mathbb{C}}

\newcommand{\N}{\mathbb{N}}

\renewcommand{\j}{\hat{\jmath}}
\renewcommand{\i}{\imath}
\renewcommand{\Re}{\mathrm{Re}}
\renewcommand{\Im}{\mathrm{Im}}

\newcommand{\ii}{\mathrm{in}}
\newcommand{\oo}{\mathrm{out}}

\newcommand{\Lt}{\mathscr{L}}

\newcommand{\lmin}{\lambda_{\mathrm{min}}}
\newcommand{\lmax}{\lambda_{\mathrm{max}}}

\newcommand{\Lb}{\mathbf{L}}
\newcommand{\Zb}{\mathbf{Z}}
\newcommand{\Yb}{\mathbf{Y}}
\newcommand{\Rb}{\mathbf{R}}
\newcommand{\Cb}{\mathbf{C}}
\newcommand{\Hi}{\mathcal{H}}

\newcommand{\Gb}{\mathbf{G}}
\newcommand{\Hr}{\mathscr{H}}
\newcommand{\Sp}{\mathbb{S}}

\DeclareMathOperator*{\spec}{spec}
\DeclareMathOperator*{\sgn}{sgn}
\DeclareMathOperator*{\dom}{dom}

\renewcommand\d{\mathop{}\!\mathrm{d}}

\newtheorem{lemma}{Lemma}
\newtheorem{corollary}{Corollary}
\newtheorem{proposition}{Proposition}
\newtheorem{theorem}{Theorem}
\newtheorem{assumption}{Assumption}

\newtheorem{solution*}{Problem}

\newtheorem{defn}{Definition}
\newtheorem{remark}{Remark}
\def\BibTeX{{\rm B\kern-.05em{\sc i\kern-.025em b}\kern-.08em
    T\kern-.1667em\lower.7ex\hbox{E}\kern-.125emX}}
\begin{document}
\title{Frequency-Domain Bounds for the Multiconductor Telegrapher's Equation}
\author{Daniel Selvaratnam, \IEEEmembership{Member, IEEE}, Alessio Moreschini, \IEEEmembership{Member, IEEE}, Amritam Das, \IEEEmembership{Member, IEEE},\\ Thomas Parisini, \IEEEmembership{Fellow, IEEE}, and Henrik Sandberg, \IEEEmembership{Fellow, IEEE}
	\thanks{Submitted for review on \today. Supported in part by the Swedish Energy Agency and ERA-Net Smart Energy Systems (project RESili8, grant agreement No 883973), the Swedish Research Council (Grant 2016-00861), and the European Union's Horizon 2020 Research and Innovation programme under grant agreement no. 739551 (KIOS CoE).}
	\thanks{D. Selvaratnam and H. Sandberg are with the Division of Decision and Control Systems, KTH Royal Institute of Technology, SE-100 44 Stockholm, Sweden. (e-mails: selv@kth.se, hsan@kth.se)}
	\thanks{A. Moreschini and T. Parisini are with the Department of Electrical and Electronic Engineering, Imperial College London, SW72AZ London, U.K. T. Parisini is also with the Department of Electronic Systems, Aalborg University, Denmark, and with the Department of Engineering and Architecture, University of Trieste, Italy (e-mails: a.moreschini@imperial.ac.uk, t.parisini@imperial.ac.uk).}
	\thanks{A. Das is with the Department of Electrical Engineering, Eindhoven University of Technology, P.O. Box 513, 5600 MB Eindhoven, The Netherlands. (e-mail: am.das@tue.nl)}}

\maketitle

\begin{abstract}
We establish mathematical bounds on the chain, ABCD and immittance matrices of a multiconductor transmission line, based on the Telegrapher's equation. Closed-form expressions for those matrices are also presented. Existing results that hold on the imaginary axis are extended to the complex plane, without reliance on \textcolor{black}{a diagonalizability assumption that is prevalent} in the literature. Therefore, the results remain valid \textcolor{black}{under arbitrary arrangements of the conducting wires, which include electrical faults. The analysis ultimately reveals important system-theoretic implications of the Telegrapher's equation, which} are of general relevance to control, power systems, and signal processing involving multiconductor transmission lines.
\end{abstract}


\section{Introduction}
\label{sec:introduction}

The Telegrapher's equation is the canonical distributed parameter model for a uniform transmission line~\cite{paul_analysis_2007}.
 From it, various two-port network parameter matrices~\cite[Chapter 2.4]{anderson_network_1973} can be derived, each of which serves as a transfer matrix between different tuples of port voltages and currents. The chain, ABCD, impedance and admittance matrices, in particular, are examined here. These transfer matrices are not rational, because they describe infinite-dimensional line dynamics. In the multiconductor case, their derivation is further complicated by the fact that matrix multiplication is not commutative. Existing expressions for the transfer matrices in the literature~\cite{paul_uniform_1973, paul_useful_1975,faria_multimodal_2014,paul_analysis_2007} restrict their domains to the imaginary axis, and \textcolor{black}{assume diagonalizability of $(\Lb \j \omega + \Rb)(\Cb \j \omega + \Gb) $ or {$(\Cb \j \omega + \Gb)(\Lb \j \omega + \Rb)$} at} all frequencies $\omega \in \R$, where $\Lb$, $\Rb$, $\Cb$, and $\Gb$ denote the inductance, resistance, capacitance, and conductance matrices of the line, respectively. 
 This assumption is widespread in the literature~\cite{faria_transfer_2020,faria_computation_2018,faria_rigorous_2017,allmeling_plecs_2024}, \textcolor{black}{and can be guaranteed under specific geometric arrangements of the conducting wires~\cite[Chapter 7.2.2]{paul_analysis_2007}. Such arrangements can break down in case of electrical faults, and scenarios that violate the assumption have been documented in~\cite{brandao_faria_overhead_1988}.} \textcolor{black}{Restriction of the domain of the network parameter matrices to the imaginary axis is also problematic, because in order to use one} for control or signal processing, knowledge of its behaviour over at least a complex right half-plane is essential.
 
 In this work, {\it closed-form expressions} for the chain, ABCD, impedance and admittance matrices of a multiconductor line are derived over a complex right-half plane, without reliance \textcolor{black}{on diagonalizability.} Growth bounds \textcolor{black}{over the right-half plane} are also established, and their \textcolor{black}{implications for network stability, causality, and transmission delays are discussed. The} results are valid for any number of conductors, and place no restrictions on the line constants beyond their fundamental physical properties. Our earlier work on fault localisation~\cite{selvaratnam_electrical_2023} concerned only a two-conductor (single phase) transmission line, and made empirical observations without establishing mathematical guarantees. It made apparent the need for a rigorous analysis of the multiconductor case, as is presented here.
 
\textcolor{black}{Treatment of multiconductor transmission lines in the literature, without assuming diagonalizability, is scant.  It is the focus of~\cite{faria_wave_1986}, which derives voltage and current solutions to the multiconductor Telegrapher's equation, and also the \emph{surge admittance} of the line, but it does not derive or analyse the chain, ABCD and immittance matrices considered here.} To the best of our knowledge, the only other work that treats multiconductor network parameter matrices with such generality is~\cite{civalleri_formal_1971}. In contrast to our work,~\cite{civalleri_formal_1971} does not present closed-form expressions, but rather establishes the existence and analyticity of the scattering and immittance matrices on the open right half-plane. It also demonstrates that the scattering matrix is bounded-real, and the immittance matrices are positive-real. Since positive-realness does not imply boundedness, the growth bounds derived herein are not mere corollaries of~\cite{civalleri_formal_1971}.
 
 After establishing basic notation and definitions below, the remainder of this paper proceeds as follows. Section \ref{sec:chain} presents the multiconductor Telegrapher's equation and, by means of the Laplace transform, derives from it the chain and ABCD matrices of the line. A blockwise decomposition of the ABCD matrix is then performed, to yield the individual A, B, C and D parameters.  Growth bounds on the chain and ABCD matrices are established, \textcolor{black}{and their physical implications discussed}. In Section \ref{sec:admittance}, the admittance matrix is constructed from the A, B, C and D parameters, and its growth \textcolor{black}{likewise bounded. The same is also} done for the impedance matrix in Section \ref{sec:impedance}, by considering a dual transmission line. The paper concludes \textcolor{black}{in Section~\ref{concl} by reviewing the results, and outlining their potential application to fault localisation.}
 
 \subsection{Preliminaries}
 \subsubsection{Sets and numbers} 
 \label{sec:sets} 
 The subset relation is denoted by $\subseteq$, and its strict version by $\subset$. 
 Let $\C$ denote the complex numbers, $\j \in \C$ the imaginary unit, $\R$ the reals, $\j \R$ the imaginary axis, $\mathbb{Z}$ the integers, and $\N = \{0,1,\hdots\}$ the naturals. Given ${\alpha \in \R}$, define $\C^+_\alpha:= \{ s \in \C \mid \Re(s)>\alpha\}$, with $\C^+:= \C^+_0$. Let $\overline{S}$ denote the closure of $S \subseteq \C$ relative to $\C$, and define ${S^+:= S \cap \C^+}$. The complex conjugate of $s \in \C$ is $\overline{s}$. 
 \subsubsection{Linear algebra}Denote the $n \times n$ identity matrix by $\I_n$. The $k$th singular value of $A \in \C^{n \times n}$ is $\sigma_k(A)$, its spectrum $\spec(A):= \{  \lambda \in \C \mid \det(A-\lambda\I_n) = 0\}$, its conjugate transpose $A^*$, and its Hermitian part $\Hr(A) :=  \frac{A + A^*}{2}$. If ${\spec (A) \subset \R}$, then $\lmin(A):= \min \spec(A)$ and $\lmax(A):= \max \spec(A)$. If $A$ is Hermitian, $A \succ 0$ asserts its positive definiteness, and $A \succeq 0$ positive semi-definiteness. The Frobenius matrix norm is denoted by $\| \cdot \|_F$.  Let $\| \cdot \|_p$ return the $p$-norm of a vector in $\C^n$, and for a matrix, the corresponding $(p,p)$-induced norm. To streamline notation, $\| \cdot \| := \| \cdot \|_2$. The numerical range of $A \in \C^{n \times n}$ is given by
 \begin{equation} W(A):= \{ x^* A x \mid x \in \Sp^n  \}, \label{eq:numrange} \end{equation}
 where the Euclidean unit sphere is
 \begin{equation} \Sp^n:= \{ x \in \C^n \mid \| x\|=1\}. \label{eq:unitSphere} \end{equation}
 \subsubsection{Functions} 
 Given a function ${f:[0,\infty) \to \R^n}$ denoted by a lower-case letter, the corresponding upper-case letter is reserved for its Laplace transform
 $$ F(s) := \Lt[f](s) : = \int_0^\infty f(t) e^{-st} \d t. $$
 Conversely, $f(t) = \Lt^{-1}[F](t) .$ 
 \begin{defn}[H-infinity] \label{def:Hinf}
 	Let $\mathcal{H}^{m \times n}_\infty$ be the space of functions $H$ such that: \begin{enumerate}
 		\item $H:S \to \C^{m \times n}$ for some $S \subseteq \C$;
 		\item $H$ is analytic on $\C^+ \subseteq S$;
 		\item $ \sup \{ \| H(s) \| \mid s \in \C^+ \} < \infty$. \label{cl:Hinf}
 	\end{enumerate}
 \end{defn}
The space $\mathcal{H}_\infty:= \bigcup_{m=1}^\infty \bigcup_{n=1}^\infty \mathcal{H}_\infty^{m \times n}$ is the set of transfer functions of causal $\mathcal{L}_2$-stable LTI systems~\cite[Chapter 3.4.3]{dullerud_course_2000}. Continuity allows the open right-half plane bound in Clause~\ref{cl:Hinf} to be extended to the imaginary axis, as per the next result. 
 \begin{proposition} \label{prop:analcont}
 	If $H \in \Hi^{m \times n}_\infty$ is continuous over ${\overline{\C^+} \subseteq \dom H}$, then $\sup \{ \| H(s) \| \mid s \in \overline{\C^+} \} < \infty$.
 \end{proposition}

\section{Chain and ABCD matrices} 
\label{sec:chain}
\subsection{Transmission line dynamics}
\label{sec:DPmodel}
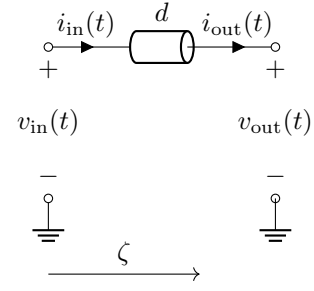
\begin{figure}[t!]
	\centering
	\begin{circuitikz}
		\draw (0,0) node[ground]{} to[open, v<={$v_\ii(t)$}, o-o] (0,2) to[short,i={$i_\ii(t)$}] (1,2) to[transmission line, label=$d$] (2,2) to [short,i={$i_\oo(t)$}] (3,2) to[open,o-o,v={$v_\oo(t)$}] (3,0) node[ground]{};
		\draw[->] (0,-1) --node[above] {$\zeta$}(2,-1);
	\end{circuitikz}
	\caption{Transmission line element of length $d>0$.} \label{fig:tline}
\end{figure}
Consider an unbroken segment of transmission line of length $d > 0$, as shown in Figure~\ref{fig:tline}. 
Let $v(t,\zeta) \in \R^n$ denote the $n$-phase line voltages, relative to ground, at time $t \geq 0$ and position $\zeta \in (0,d)$ along the line. Let $i(t,\zeta) \in \R^n$ denote the corresponding line currents, where current in the direction of increasing $\zeta$ is defined as positive. The relationship between currents and voltages along the interior of the line is modelled by the Telegrapher's equation~\cite[(3.27)]{paul_analysis_2007}: $\forall t > 0,\ \zeta \in (0,d),$
\begin{equation}\frac{\partial x}{\partial \zeta}(t,\zeta) =- \begin{bmatrix}
		0 & \Lb \\ \Cb & 0
	\end{bmatrix}\frac{\partial x}{\partial t}(t,\zeta) - \begin{bmatrix}
		0 & \Rb \\ \Gb & 0
	\end{bmatrix} x(t,\zeta), \label{eq:stateDynamics} \end{equation}
where ${x(t,\zeta) := \begin{bsmallmatrix}
		v(t,\zeta)\\i(t,\zeta)
\end{bsmallmatrix}}$, and for a uniform line, $\Lb,\Cb,\Rb,\Gb \in \R^{n \times n}$ are known constants.
\smallskip
\begin{assumption}[Constants] \label{ass:parameters}
	The matrices ${\Lb,\Cb,\Rb,\Gb \in \R^{n \times n}}$ are symmetric, with ${\Lb,\Cb \succ 0}$.
\end{assumption}
\smallskip
The diagonal elements of $\Lb,\Cb,\Rb$ and $\Gb$ correspond, respectively, to the inductance, capacitance, line-resistance and conductance-to-ground of each phase, per unit length. The off-diagonal terms result from electromagnetic interactions between different phases, as well as the external environment. 
\smallskip
\begin{remark}[Physical properties]
	The symmetry of $\Lb, \Cb, \Rb$ and $\Gb$ is established in \cite[Chapter 3]{paul_analysis_2007} from first principles. Positive definiteness of $\Lb$ and $\Cb$ is also established in \cite[Section 3.5]{paul_analysis_2007}.
\end{remark}
\smallskip
Finally, the boundary conditions
	\begin{align}
		& \lim_{\zeta  \to 0^+} x(t,\zeta) = \begin{bmatrix} v_\ii(t) \\ i_\ii(t) \end{bmatrix}, &&  \lim_{\zeta  \to d^-} x(t,\zeta) = \begin{bmatrix} v_\oo(t)\\i_\oo(t) \end{bmatrix},	\label{eq:BCs}
	\end{align}
prescribe the voltages $v_\ii(t), v_\oo(t)\in \R^n$ and currents $i_\ii(t), i_\oo(t)\in \R^n$ at the two ends of the line. 

\subsection{Chain matrix derivation}
To derive the transfer function from $\begin{bsmallmatrix}
	v_\ii \\ i_\ii
\end{bsmallmatrix}$ to $\begin{bsmallmatrix}
	v_\oo \\ i_\oo
\end{bsmallmatrix}$, consider the following solution space. 
\begin{defn}[Solution space] \label{def:soln}
	Let $\mathcal{X}$ be the set of continuously differentiable functions $x:\R^+ \times (0,d) \to \R^{2n}$ that satisfy all the following:
	\begin{enumerate}
		\item initial rest: \label{cl:rest} $	\forall \zeta \in (0,d),\ \lim_{t \to 0^+} x(t,\zeta) = 0 ;$
		\item exponential order in time: \label{cl:exporeder} there exists $\alpha \in \R$ such that
		\begin{align*}  
			 \forall \zeta \in (0,d),\ &\exists M,T \geq 0,\\ \forall t \geq T,\
			 &\| x(t,\zeta)\|,  \| \tfrac{ \partial x}{\partial t}(t,\zeta)\| \leq Me^{\alpha t};
		\end{align*}
		\item interchange of temporal Laplace transformation and spatial differentiation: \label{cl:diff} for all $s \in \C^+_\alpha$ and $\zeta \in (0,d)$,
		$$ \int_0^\infty \frac{\partial x}{\partial \zeta} (t,\zeta) e^{-st} \d t =\frac{\partial}{\partial \zeta} \int_0^\infty x(t,\zeta) e^{-st} \d t;$$
		\item interchange of Laplace transformation and limits: \label{cl:lim} for every $s \in \C^+_\alpha$ and $p \in \{0^+,d^-\}$,
$$ \textcolor{black}{ \lim_{\zeta \to p} \int_0^\infty x(t,\zeta) e^{-st} \d t =  \int_0^\infty \lim_{\zeta \to p} x(t,\zeta) e^{-st} \d t. }$$
	\end{enumerate}
\end{defn}
\smallskip
These restrictions on the solution space permit derivation of the transfer function according to \cite[Section 39]{doetsch_introduction_1974}.
Clauses~\ref{cl:exporeder}--\ref{cl:lim} of Definition \ref{def:soln}, in particular, formalise assumptions~\cite[$W_1$--$W_3$, Section 39.1]{doetsch_introduction_1974}.

Suppose there exists a solution $x \in \mathcal{X}$ to~\eqref{eq:stateDynamics} that satisfies the boundary conditions~\eqref{eq:BCs}. 
At each $\zeta \in (0,d)$, the Laplace transform
$$ X(s,\zeta):= \int_0^\infty x(t,\zeta) e^{-st} \d t$$
converges on $\C_\alpha^+$, by Clause \ref{cl:exporeder}. The same holds for the Laplace transform of $\frac{\partial x}{\partial t}$.
Take Laplace transforms of both sides of \eqref{eq:stateDynamics} to obtain
\begin{align} \frac{\partial X}{\partial \zeta}(s,\zeta) = -\begin{bmatrix}
		0 & \Lb s + \Rb \\ \Cb s + \Gb & 0
	\end{bmatrix} X(s,\zeta),\label{eq:ODE} \end{align}
\textcolor{black}{under Clauses \ref{cl:rest}--\ref{cl:diff} of Definition \ref{def:soln}.} Since \eqref{eq:ODE} is now a linear ODE in $\zeta$, its solution satisfies
\begin{equation*} \forall \zeta,\zeta_0 \in (0,d)\ \forall s \in \C^+_\alpha,\quad X(s,\zeta) =  \Xi(s, \zeta_0-\zeta) X(s,\zeta_0), \end{equation*}
where $\Xi:\C \times \R \to \C^{2n \times 2n}$ is the matrix exponential
\begin{equation} \Xi(s,\zeta) := e^{\zeta\begin{bmatrix}
			0 & \Lb s + \Rb \\ \Cb s + \Gb & 0
	\end{bmatrix}}. \label{eq:Xi} \end{equation}
Taking $\zeta_0 \to 0$ and $\zeta \to d$, Clause \ref{cl:lim} gives
\begin{equation} \begin{bsmallmatrix}
		V_\oo(s) \\ I_\oo(s)
	\end{bsmallmatrix}  = \Xi(s,-d) \begin{bsmallmatrix}
		V_\ii(s) \\ I_\ii(s)
	\end{bsmallmatrix}. \label{eq:TF} \end{equation}
Thus, the mapping $s \mapsto \Xi(s,-d)$ is the desired transfer function, with $\Xi(s,-d)$ being the \emph{chain matrix}~\cite[Section 6.5.1]{paul_analysis_2007} of a transmission line of length $d$. 
\begin{remark}[Port conventions] \label{rem:port_conventions}
\textcolor{black}{Given \eqref{eq:stateDynamics} and the direction of the $\zeta$-coordinate, the port labels in Figure~\ref{fig:tline} are chosen such that the transfer function from input port to output port is the chain matrix. Likewise, every other network parameter matrix corresponds to a different designation of inputs and outputs.}
\end{remark}
\subsection{ABCD matrix}
From \eqref{eq:Xi}, the first observation is that $\Xi(s,d) = \Xi(s,-d)^{-1}$, which is the inverse of the chain matrix. It thereby satisfies,
\begin{equation} \begin{bsmallmatrix}
		V_\ii(s) \\ I_\ii(s)
	\end{bsmallmatrix} = \Xi(s,d) \begin{bsmallmatrix}
		V_\oo(s) \\ I_\oo(s)
	\end{bsmallmatrix}, \label{eq:ABCDrel} \end{equation}
making it the \emph{ABCD matrix}~\cite{frickey_conversions_1994} of the line. The following result demonstrates that the chain and ABCD matrices are unitarily similar.
\begin{proposition} \label{prop:norm}
	For any $s \in \C$ and $d \in \R$, $$\Xi(s,d) = \begin{bsmallmatrix}
		-\I_n & 0 \\ 0 & \I_n
	\end{bsmallmatrix}^* \Xi(s,-d) \begin{bsmallmatrix}
	-\I_n & 0 \\ 0 & \I_n
	\end{bsmallmatrix},$$
	and thus $\| \Xi(s,d) \| = \| \Xi(s,-d) \|$.
\end{proposition}
\begin{proof}
Let $M := \begin{bsmallmatrix}
	0 & \Lb s + \Rb \\ \Cb s + \Gb & 0
\end{bsmallmatrix}$ and $U := \begin{bsmallmatrix}
-\I_n & 0 \\ 0 & \I_n
\end{bsmallmatrix}$. Then, $-M = UMU^*$. Thus, 
$\Xi(s,-d) =  e^{-dM} = e^{UdMU^*} = Ue^{dM}U^* = U\Xi(s,d)U^*$.
\end{proof}
In light of this, properties derived for the ABCD matrix can be applied directly to the chain matrix, and \emph{vice versa}. 

\subsection{Blockwise decomposition of ABCD matrix}
In order to derive the other network parameter matrices, the ABCD matrix must be partitioned into four square blocks, which are themselves the A, B, C and D network parameters.

\smallskip
\begin{proposition} \label{prop:BlockAntidiagonalExponential} 
	Let $A,B\in \C^{n \times n}$. If there exist invertible $P,Q \in \C^{n \times n}$ such that $P^2 = AB$ and $Q^2 = BA$, then
	$$ e^{\begin{bsmallmatrix}
			0 & A \\ B & 0
	\end{bsmallmatrix}} = 
	\begin{bmatrix}
		\cosh P & AQ^{-1} \sinh Q \\ BP ^{-1}\sinh P & \cosh Q
	\end{bmatrix}.$$
\end{proposition}
\begin{proof}
	Let $ M := \begin{bsmallmatrix}
		0 & A \\ B & 0
	\end{bsmallmatrix} $. The matrix exponential, $
		e^M = \sum_{k=0}^\infty \frac{M^k}{k!}$, converges for all square matrices. 
	Observe that \begin{align} M^{2k} &= \begin{bmatrix}
			(AB)^k & 0 \\ 0 & (BA)^k
		\end{bmatrix}, \label{eq:evenPowers} \\
		M^{2k+1} &= \begin{bmatrix}
			0 & A(BA)^k \\ B(AB)^k & 0
		\end{bmatrix}, \label{eq:oddPowers}
	\end{align} for any $k \in \N$. Hence, splitting the summation into even and odd powers gives
	\begin{align*}
		e^M & = \sum_{k=0}^\infty \left( \frac{M^{2k}}{(2k)!} + \frac{M^{2k+1}}{(2k+1)!} \right) \\
		& =  \medmath{ \sum_{k=0}^\infty \left( \frac{\begin{bmatrix}
					(AB)^k & 0 \\ 0 & (BA)^k
			\end{bmatrix} }{(2k)!}+  \frac{\begin{bmatrix}
					0 & A(BA)^k \\ B(AB)^k & 0
			\end{bmatrix} }{(2k+1)!}\right)} \\
		& =  \sum_{k=0}^\infty \begin{bmatrix}
			\frac{(AB)^k }{(2k)!} & \frac{A(BA)^k }{(2k+1)!} \\
			\frac{B(AB)^k }{(2k+1)!} & \frac{(BA)^k }{(2k)!}
		\end{bmatrix} \,.
	\end{align*} 
	Suppose $P,Q \in \C^{n \times n}$ are such that $P^2 = AB$ and $Q^2 = BA$.
	The matrix hyperbolic functions are given by the globally convergent~\cite[Section 4.3]{higham_functions_2008} power series \begin{align*}
		\cosh P &= \sum_{k=0}^\infty \frac{P^{2k}}{(2k)!} = \sum_{k=0}^\infty \frac{(AB)^k}{(2k)!}, \\
		\cosh Q &= \sum_{k=0}^\infty \frac{Q^{2k}}{(2k)!} = \sum_{k=0}^\infty \frac{(BA)^k}{(2k)!}, \\
		\sinh P &= \sum_{k=0}^\infty \frac{P^{2k+1}}{(2k+1)!} = P\sum_{k=0}^\infty \frac{(AB)^k}{(2k+1)!},\\
		\sinh Q &= \sum_{k=0}^\infty \frac{Q^{2k+1}}{(2k+1)!} = Q\sum_{k=0}^\infty \frac{(BA)^k}{(2k+1)!}.
	\end{align*}
	If $P,Q$ are also invertible, then the result follows. \end{proof}
\smallskip
Clearly, before Proposition~\ref{prop:BlockAntidiagonalExponential} can be applied to $\Xi(s,d)$, invertible matrix square roots of $(\Lb s + \Rb)(\Cb s + \Gb)$ and $(\Cb s + \Gb)(\Lb s + \Rb)$ must be found.
\smallskip
\begin{defn}[Principal square root] \label{def:root}
	A \emph{principal square root} of ${A \in \C^{n \times n}}$ is a matrix $X \in \C^{n \times n}$ such that both ${\spec(X) \subset \C^+}$ and $X^2 = A$. When such an $X$ is unique, denote it by $\sqrt{A}$. 
\end{defn}
\smallskip

Note that a principal square root is always invertible, because its spectrum cannot contain 0. Now a square root of a Hermitian matrix is easily obtainable via diagonalization, but even though $\Lb, \Cb,\Rb,\Gb$  are symmetric, $\Lb s + \Rb$ and $\Cb s + \Gb$ are not Hermitian when $s \notin \R$. They do, however, fall within another convenient class of matrices.

\begin{defn}[Accretive matrix] \label{def:accretive} 
	A square complex matrix $M$ is \emph{accretive} if its Hermitian part $ \Hr(M) \succ 0$.
\end{defn}

\begin{proposition}
\label{prop:sqrts}
	If $A,B \in \C^{n \times n}$ are accretive, then the unique principal square root $\sqrt{AB} $ exist. 
\end{proposition}
\begin{proof} 
	Any real eigenvalues of a product of accretive matrices are strictly positive~\cite[Theorem 1]{ballantine_accretive_1975}. Thus, $\spec(AB) \cap (-\infty,0] = \emptyset$. Since $AB$ has no eigenvalues in $(-\infty,0]$, it has a unique principal square root~\cite[Theorem 1.29]{higham_functions_2008}.
	\end{proof}
\smallskip

To see that $\Lb s + \Rb$ and $\Cb s + \Gb$ are both accretive over a right half-plane, first consider the linear combination of two Hermitian matrices.
\smallskip
\begin{lemma} \label{lem:scaling}
	If $A,B \in \C^{n \times n}$ are Hermitian and $t  \in \R$, then 
\begin{align*}  
	&\lmin(At) + \lmin(B)  \leq  \lmin(At + B) .
\end{align*}
\end{lemma}
\begin{proof} Recalling that $\Sp^n$ is the unit sphere by \eqref{eq:unitSphere}, we have
	\begin{align*}  
		\lmin(At + B) & = \min \{ x^* (At + B)x \mid x \in \Sp^n \} \\
		& \medmath{\geq \min \{ x^*Atx \mid x \in \Sp^n \} + \min \{ x^*Bx \mid x \in \Sp^n \} } \\
		& = \lmin(At) + \lmin(B).
	\end{align*}
	\end{proof}

This lower bound is now used to show that the sum of a positive definite matrix with an arbitrary Hermitian matrix remains positive definite, if the former is sufficiently large.

\begin{lemma} \label{lem:posdef}
	Let $A,B \in \C^{n \times n}$ be Hermitian. If $A \succ 0$, then $At + B \succ 0$ for all $ t > -\min  \left\{ \frac{\lmin(B)}{ \lmax(A)} ,  \frac{\lmin(B)}{ \lmin(A)}\right \}.$
\end{lemma}
\begin{proof} 
	Let $t> \max \left\{ -\frac{\lmin(B)}{ \lmax(A)} ,  -\frac{\lmin(B)}{ \lmin(A)} \right\}.$
	Lemma \ref{lem:scaling} implies
	\begin{equation*} \lmin(A t + B) \geq \lmin(At) + \lmin(B). \end{equation*}
	Suppose $A \succ 0$, by which $\lmax(A) \geq \lmin(A) > 0$. 
	If ${t \geq 0}$, then $\lmin(At) = t\lmin(A) > - \lmin(B)$, and it follows that $\lmin(At + B) > 0$. 
	Otherwise, $\lmin(At) = t \lmax(A) > - \lmin(B),$ which yields the same result. 
	\end{proof}
\smallskip

These results for Hermitian matrices can now be applied to $\Lb s + \Rb$ and $\Cb s + \Gb$ by taking their Hermitian parts.

\begin{corollary}  \label{cor:realparts}
	Under Assumption~\ref{ass:parameters}, $\Lb s + \Rb$ is accretive for all $s \in \C^+_\rho$, and $\Cb s + \Gb$ is accretive for all $s \in \C^+_\gamma$, 
	where 
	\begin{subequations}
		\begin{align} \gamma&:=-\min\left\{\tfrac{\lmin(\Gb)}{\lmax(\Cb)},\tfrac{\lmin(\Gb)}{\lmin(\Cb)} \right\}, \label{eq:gamma} \\ \rho&:=-\min\left\{\tfrac{\lmin(\Rb)}{\lmax(\Lb)},\tfrac{\lmin(\Rb)}{\lmin(\Lb)} \right\}.  \label{eq:rho} \end{align}
		\label{eq:rhogamma}
	\end{subequations}
\end{corollary}
\begin{proof} For any $s \in \C$, $\Hr(\Lb s + \Rb)  = \Lb \Re(s) + \Rb  $ is symmetric, and if $\Re(s) > \rho$, is also positive definite by Lemma~\ref{lem:posdef}. Similarly, if $\Re(s)>\gamma$, then $\Hr(\Cb s + \Gb) = \Cb \Re(s) + \Gb \succ 0$. \end{proof}
\smallskip

In contrast to \cite[Chapter 7]{paul_analysis_2007}, which relies \textcolor{black}{on diagonalizability} to construct the square roots, their existence can now simply be guaranteed by changing the domain of evaluation from the imaginary axis to a right half-plane.

\begin{corollary} \label{cor:PSR}
	Under Assumption~\ref{ass:parameters}, for any ${s \in \C^+_\rho \cap \C^+_\gamma}$, the unique principal square roots $\sqrt{(\Lb s + \Rb)(\Cb s + \Gb)}$ and $\sqrt{(\Cb s + \Gb)(\Lb s + \Rb)}$ both exist.
\end{corollary}

Piecing together the preceding results yields the desired blockwise decomposition of the ABCD matrix.
\smallskip
\begin{theorem}[ABCD matrix] \label{thm:CPM}
	Under Assumption~\ref{ass:parameters},
	\begin{align*} 
		\Xi(s,d) := e^{d\begin{bsmallmatrix}
				0 & \Lb s + \Rb \\ \Cb s + \Gb & 0
		\end{bsmallmatrix}} = \begin{bmatrix}
			A_d(s) & B_d(s) \\ C_d(s) & D_d(s)
	\end{bmatrix} \end{align*}
	for any $s \in \C^+_\alpha$ and $d \in \R$, where
	\begin{subequations}
		\begin{align}
			\alpha&:= \medmath{ -\min\left\{\tfrac{\lmin(\Gb)}{\lmax(\Cb)},\tfrac{\lmin(\Gb)}{\lmin(\Cb)},\tfrac{\lmin(\Rb)}{\lmax(\Lb)},\tfrac{\lmin(\Rb)}{\lmin(\Lb)} \right\},} \label{eq:alpha} \\
			\Zb(s)& := (\Lb s + \Rb) \sqrt{(\Cb s + \Gb)(\Lb s + \Rb)}^{-1},  \label{eq:Z1} \\
			\Yb(s) &:= (\Cb s + \Gb) \sqrt{(\Lb s + \Rb)(\Cb s + \Gb)}^{-1}, \label{eq:Z2} \\
			A_d(s)  &:= \cosh \left( d \sqrt{(\Lb s + \Rb)(\Cb s + \Gb)}\right),\\
			B_d(s) & := \Zb(s) \sinh\left(d \sqrt{(\Cb s + \Gb)(\Lb s + \Rb)} \right), \label{eq:Bd} \\
			C_d(s) & := \Yb(s) \sinh\left(d \sqrt{(\Lb s + \Rb)(\Cb s + \Gb)} \right), \\
			D_d(s) & := \cosh \left( d \sqrt{(\Cb s + \Gb)(\Lb s + \Rb)} \right). \label{eq:Dd}
		\end{align} 
        \label{eq:ABCD}
	\end{subequations}
\end{theorem}
\begin{proof} 
	Observe that $\C^+_\rho \cap \C^+_\gamma = \C^+_\alpha$, since
	$ \alpha = \max\{\rho,\gamma \} $.
	The rest is a direct consequence of Proposition~\ref{prop:BlockAntidiagonalExponential} and Corollary~\ref{cor:PSR}.  \end{proof}
\smallskip

	The above blockwise decomposition is valid on $\C^+_\alpha$, which contains the imaginary axis when $\Gb,\Rb \succ 0$. In this case, Theorem \ref{thm:CPM} directly generalises the various results in \cite[Section 7.2]{paul_analysis_2007}, which evaluate $\Xi(s,-d)$ along the imaginary axis under assumptions that render $(\Lb \j  \omega + \Rb)(\Cb \j  \omega + \Gb)$ and $(\Cb \j  \omega + \Gb)(\Lb \j  \omega + \Rb)$ \textcolor{black}{diagonalizable. (Since both factors are symmetric, if one of these products is diagonalizable, then the other is too.)}
	
\subsection{Growth bounds}
Bounding the chain and ABCD matrices over the whole complex plane 
first involves bounding the exponent $ \begin{bsmallmatrix}
	0 & \Lb  s + \Rb \\ \Cb s + \Gb
\end{bsmallmatrix} $ along the real axis. 

\begin{lemma} \label{lem:Delta}
Under Assumption \ref{ass:parameters},
\begin{equation}
	\begin{Vmatrix}
		0 & \Lb  \varsigma + \Rb \\ \Cb \varsigma + \Gb
	\end{Vmatrix} \leq c_1 | \varsigma| + c_0
\end{equation}
for all $\varsigma \in \R$, where 
\begin{subequations}
	\begin{align}
		c_0 &:= \max \{ \| \Rb\|,  \| \Gb \| \} \geq 0, \\
		c_1 &:= \max \{ \|\Lb \|, \| \Cb \| \} >0 .
	\end{align}
	\label{eq:c}
\end{subequations}
\end{lemma}
\begin{proof}
Applying \cite[Lemma 2.10]{zhou_robust_1996},
\begin{align*}
	&\begin{Vmatrix}
		0 & \Lb  \varsigma + \Rb \\ \Cb \varsigma + \Gb
	\end{Vmatrix}  \leq \begin{Vmatrix}
	0 & \| \Lb  \varsigma + \Rb \|  \\  \| \Cb \varsigma + \Gb \|
	\end{Vmatrix} \\
	& \leq \begin{Vmatrix}
		0 & \|\Lb\| |\varsigma|  +  \|\Rb\|   \\ \|\Cb\| |\varsigma| + \| \Gb \| &  0
	\end{Vmatrix}.
\end{align*}
Since $\begin{Vsmallmatrix} 0 & a \\ b & 0\end{Vsmallmatrix} \leq \max\{a,b\}$ for $a,b \geq 0$, the result follows.
\end{proof}
To bound the magnitude of the ABCD matrix along the imaginary axis, temporarily consider the lossless case ($\Rb = \Gb = 0$).
\begin{lemma} \label{lem:lossless}
	Under Assumption \ref{ass:parameters},
	\begin{equation}  \kappa:= \sup \left\{  \left \| \exp \begin{bmatrix}
		0 &  \j \omega \Lb \\ \j \omega \Cb  & 0
	\end{bmatrix} \right\| \mid \omega \in \R \right\} \in [1,\infty). \label{eq:kappa} \end{equation}
If, moreover, $\Cb\Lb$ is normal, then  \begin{equation} \kappa \leq 1 + \frac{\max \{ \|\Lb \|, \| \Cb \| \}}{\sqrt{\lmin(\Cb\Lb)}} . \label{eq:kbound}  \end{equation}
\end{lemma}
\begin{proof}
	To see that $\kappa \geq 1$, consider $\omega = 0$. 
	
	Its finiteness is established next. The products $\Cb \Lb$ and $\Lb \Cb$ are both diagonalizable, and have strictly positive eigenvalues by \cite[Corollary 7.6.2]{horn_matrix_2012}. Therefore, they have unique principal square roots by \cite[Theorem 1.29]{higham_functions_2008}.
	In particular, there exists an invertible $Q \in \C^{n \times n}$ such that 
	\begin{equation} \Lambda:= Q^{-1} \Lb \Cb Q  = Q^* \Cb  \Lb Q^{-*} \succ 0 \label{eq:W} \end{equation}
	is real and diagonal, by which \begin{align} &\sqrt{\Lb \Cb} = Q \sqrt{\Lambda} Q^{-1}, &\sqrt{\Cb \Lb} = Q^{-*} \sqrt{\Lambda}  Q^*. \label{eq:sqrtCL} \end{align} Applying Proposition \ref{prop:BlockAntidiagonalExponential}, $\exp \left( \begin{bsmallmatrix}
		0 &  \j \omega \Lb \\ \j \omega \Cb  & 0
	\end{bsmallmatrix} \right) = $ 
	\begin{align*} 
		&\begin{bmatrix}
			\cosh \left( \j \omega \sqrt{ \Lb \Cb} \right) & \Lb\sqrt{\Cb\Lb} ^{-1} \sinh \left( \j \omega \sqrt{\Cb\Lb} \right)   \\ \Cb \sqrt{ \Lb \Cb} ^{-1}\sinh \left( \j \omega \sqrt{ \Lb \Cb} \right) & \cosh \left( \j \omega \sqrt{\Cb\Lb} \right) 
		\end{bmatrix} =\\
		&\begin{bmatrix}
			\cos \left( \omega \sqrt{ \Lb \Cb} \right) & \j \Lb\sqrt{\Cb\Lb} ^{-1} \sin \left( \omega \sqrt{\Cb\Lb} \right)   \\ \j \Cb \sqrt{ \Lb \Cb} ^{-1}\sin \left( \omega \sqrt{ \Lb \Cb} \right) & \cos \left( \omega \sqrt{\Cb\Lb} \right) 
		\end{bmatrix} =\\
		& \medmath{ \begin{bmatrix}
				Q \cos \left( \omega \sqrt{ \Lambda} \right) Q^{-1}& \j \Lb\sqrt{\Cb\Lb} ^{-1} Q^{-*}\sin \left( \omega \sqrt{\Lambda} \right)Q^*   \\ \j \Cb \sqrt{ \Lb \Cb} ^{-1}Q \sin \left( \omega \sqrt{ \Lambda} \right)Q^{-1} & Q^{-*} \cos \left( \omega \sqrt{\Lambda} \right) Q^*\\
		\end{bmatrix}}.
	\end{align*}
Since every component of the above matrix is bounded, ${\kappa < \infty}$. 
	
	If $\Cb\Lb$ is normal, then a unitary $Q$ can be chosen in \eqref{eq:W}, in which case $Q^* = Q^{-1}$ and $Q^{-*} = Q$. Now $\| \sin(\omega \sqrt\Lambda)\|, \| \cos(\omega \sqrt\Lambda)\| \leq 1$ for all $\omega$. Since the spectral norm is unitarily invariant, \cite[Lemma 2.10]{zhou_robust_1996} yields
	\begin{align*}
		\left\| \exp \left( \begin{bsmallmatrix}
			0 &  \j \omega \Lb \\ \j \omega \Cb  & 0
		\end{bsmallmatrix} \right) \right\| &\leq \begin{Vmatrix}
				1 & \| \Lb\sqrt{\Cb\Lb} ^{-1} \|   \\ \| \Cb \sqrt{ \Lb \Cb} ^{-1}\|  & 1\\
		\end{Vmatrix}.
		\end{align*}
		It is also clear from \eqref{eq:sqrtCL} that
		$$ \| \sqrt{\Lb\Cb}^{-1}\|  = \| \sqrt{\Cb\Lb}^{-1}\|   =  \lmin(\Cb\Lb)^{-\frac{1}{2}}.$$
	Recalling $\begin{Vsmallmatrix} 0 & a \\ b & 0\end{Vsmallmatrix} \leq \max\{a,b\}$ for $a,b \geq 0$, \eqref{eq:kbound} follows.
\end{proof}
Combining the preceding bounds yields an exponential bound on $\Xi(s,d)$ over the whole complex plane, in full generality. 
\begin{theorem}[Growth of ABCD matrix] \label{thm:growth}
	Under Assumption \ref{ass:parameters},
	\begin{equation} \| \Xi(s,d) \| \leq \kappa e^{\frac{|d|}{\nu}( |\Re (s)| + \theta)} \label{eq:expbound} \end{equation}
	for all $s \in \C$ and $d \in \R$, where
\begin{subequations}
	\begin{align}
		\theta&:= \frac{ \max\{ \|\Rb\|,\| \Gb\| \}}{\max\{ \| \Lb \|, \| \Cb \| \}} \geq 0,\\
		\nu&:=  \frac{1}{\kappa \max\{ \| \Lb \|, \| \Cb \| \}} > 0.
	\end{align}
	\label{eq:params}
\end{subequations}
\end{theorem}
\begin{proof}
	For any $d,\varsigma,\omega \in \R$, 
	\begin{align*}
		\Xi(\varsigma + \j \omega,d) = e^{(A + \Delta)|d|},
	\end{align*}
	where $A :=  \j \omega \sgn(d) \begin{bsmallmatrix}
		0 & \Lb \\ \Cb & 0
	\end{bsmallmatrix}$ and $\Delta := \sgn(d) \begin{bsmallmatrix}
	0 & \Lb \varsigma + \Rb \\ \Cb \varsigma + \Gb & 0
	\end{bsmallmatrix} $. By Lemma \ref{lem:lossless},  $\| e^{A|d|}\| \leq \kappa < \infty$ for all $d \in \R$.  It follows from \cite[Proposition 4.2.18]{hinrichsen_mathematical_2005} that $\|e^{(A + \Delta)d} \| \leq \kappa e^{\kappa \|\Delta \| |d|}$, where $ \| \Delta \| \leq c_1 |\zeta| + c_0$ by Lemma~\ref{lem:Delta}. The exponent is thus $ \kappa c_1(|\zeta| + \frac{c_0}{c_1})|d|$, and the result follows from \eqref{eq:c}. 
\end{proof}
\smallskip
\subsection{Implications of bounds}
\label{sec:implications}
\textcolor{black}{
It is worth examining the physical implications of Theorem~\ref{thm:growth}. 
}
\subsubsection{Stability}
\textcolor{black}{
The bound \eqref{eq:expbound} is independent of the imaginary part of $s$. Thus, the chain and ABCD matrices are bounded along every vertical line in the complex plane, including the imaginary axis.
\begin{corollary}[$\mathcal{L}_2$-Stability] \label{cor:imagAxis}
	Under Assumption~\ref{ass:parameters},
	$$ \forall d \in \R,\ \sup \{ \| \Xi(\j \omega,d) \| \mid \omega \in \R \} < \infty.$$ 
\end{corollary}
In other words, both chain and ABCD matrices are elements of $\mathcal{L}_\infty(\j \R)$, which establishes $\mathcal{L}_2$-stability of their corresponding time-domain operators~\cite[Theorem 3.25]{dullerud_course_2000}. For the chain matrix, this guarantees the energy at the output port in Figure~\ref{fig:tline} is bounded linearly by the energy at the input port. A linear bound in the opposite direction follows from the $\mathcal{L}_2$-stability of the ABCD matrix.}

\subsubsection{Causality} 
\textcolor{black}{The bound \eqref{eq:expbound} grows exponentially with the real part of $s$. Whilst it is only an upper bound, consideration of the two-conductor ($n=1$) case reveals that the chain and ABCD matrices do indeed grow exponentially as ${\Re(s) \to \infty}$. Hence, these transfer functions are not in $\mathcal{H}_\infty$, and are therefore not causal~\cite[Chapter 3.4.3]{dullerud_course_2000}. }

\textcolor{black}{
To understand the implications of this, consider two identical transmission lines. If the input signals to each are equal up to some time $t$, network acausality means that the outputs may differ even before time $t$. Differences in the input after time $t$ can thus ``travel back in time" to affect the output before time $t$. Of course, tranmission lines do not really permit electrons to travel back in time (to the best of the authors' knowledge). The mathematical fact that these transfer matrices are acausal indicates that the signals labelled ``inputs" should not be viewed as ``causing" the signals labelled ``outputs". Rather, both inputs and outputs satisfy a joint constraint in their product signal space~\cite{willems_behavioral_2007}, and information flows both ways. If the propagation of information were instantaneous, then causality would still be preserved. However, since the speed of propagation is finite, disturbances at an ``output" terminal take time to reach an ``input" terminal, so that the output is seen to change before the input. Thus, the acausality here reflects the physical propagation delays in the transmission line.  In fact, we now show that the rate of exponential growth in \eqref{eq:expbound} yields a lower-bound on the propagation speed.}

\begin{remark}[Lead-time] \label{rem:lead}
	For any fixed $d \in \R$, an ABCD matrix can be written as the product $$\Xi(s,d) = H_d(s) e^{ \frac{|d|}{\nu}s}$$ of a pure time advance with the causal transfer function 
	$$ H_d(s) := e^{-\frac{|d|}{\nu}s}\Xi(s,d),$$  
	where $H_d \in \mathcal{H}_\infty$ by Theorem \ref{thm:growth}. The \textcolor{black}{exponent $\frac{|d|}{\nu} $ in \eqref{eq:expbound} thus} bounds the time by which the output can lead the input, \textcolor{black}{thereby quantifying the degree of acausality in the system. This, in turn, makes} $\nu$ a lower-bound on the speed of information propagation. The same \textcolor{black}{relationships} clearly hold for the chain parameter matrix, \textcolor{black}{by} Proposition \ref{prop:norm}. 
\end{remark}
\section{The admittance matrix} \label{sec:admittance}
\subsection{Derivation}
In contrast to \eqref{eq:TF} and \eqref{eq:ABCDrel}, the admittance matrix $Y(s,d)$ must satisfy
\begin{equation}
	\begin{bsmallmatrix}
		I_\ii(s) \\ -I_\oo(s)
	\end{bsmallmatrix} =  Y(s,d) \begin{bsmallmatrix}
		V_\ii(s) \\ V_\oo(s)
	\end{bsmallmatrix} \label{eq:admittanceRel}
\end{equation}
by definition, with $-i_\oo$ directed into the line as required. Applying the two-port conversion formula \cite[(13)]{reveyrand_multiport_2018},
\begin{equation}
	Y(s,d) = \begin{bmatrix}
		D_d(s)B_d(s)^{-1} & C_d(s) - D_d(s)B_d(s)^{-1}A_d(s) \\
		-B_d(s)^{-1} & B_d(s)^{-1}A_d(s)
	\end{bmatrix}. \label{eq:primalAdmittance}
\end{equation}
To simplify this, observe
$$ \medmath{ \Xi(s, -d) \Xi(s,d) = \begin{bsmallmatrix}
		A_d(s) & -B_d(s) \\ -C_d(s) & D_d(s)
	\end{bsmallmatrix} \begin{bsmallmatrix}
		A_d(s) & B_d(s) \\ C_d(s) & D_d(s)
	\end{bsmallmatrix}  = \begin{bsmallmatrix}
		\I_n & 0 \\ 0 & \I_n
	\end{bsmallmatrix},}$$
by which $0 = A_d(s)B_d(s) - B_d(s) D_d(s) $, so
$ B_d(s)^{-1} A_d(s) = D_d(s) B_d(s)^{-1}$. Additionally, $ \I_n = -C_d(s) B_d(s) + D_d(s)^2 $,
so $B_d(s)^{-1} = -C_d(s)  + D_d(s)^2 B_d(s)^{-1} =  -C_d(s) + D_d(s) B_d(s)^{-1}A_d(s)$.
Thus, the line admittance matrix is given by
\begin{equation}
	Y(s,d) = \begin{bmatrix}
		D_d(s)B_d(s)^{-1} & -B_d(s)^{-1} \\
		-B_d(s)^{-1} & D_d(s)B_d(s)^{-1}
	\end{bmatrix}. \label{eq:admittance}
\end{equation}
The invertibility of $B_d(s)$ is verified in the next two results

\smallskip
\begin{lemma}
	For any $X \in \C^{n \times n}$, $\sinh(X)$ is singular if and only if $\spec(X) \cap \{  k \pi \j \mid k \in \mathbb{Z} \} \neq \emptyset.$
\end{lemma}
\begin{proof} 
	Suppose $\sinh(X)$ is singular. Then there exists a nonzero $z \in \C^n$ such that $\sinh(X)z = \left( \frac{e^X - e^{-X}}{2}\right)z = 0$. It follows that $e^Xz = e^{-X}z$, by which $ e^{2X}z = z.$
	Thus, 1 is an eigenvalue of $e^{2X}$. Since $\spec(e^{2X}) = \{ e^ \lambda \mid \lambda \in \spec(2X)\}$ by ~\cite[Proposition 11.2.3]{bernstein_matrix_2009}, there exists $\lambda \in \spec(2X)$ such that $e^\lambda = 1$. It follows that $\lambda \in \{ 2 k \pi \j \mid k \in \mathbb{Z}  \}$. Thus, $\frac{\lambda}{2} \in \spec(X) \cap \{  k \pi \j \mid k \in \mathbb{Z} \} $. 
	To establish the converse, retrace the same argument in the opposite direction. \end{proof}
\smallskip
\begin{corollary} \label{cor:Bdinv}
	Under the hypotheses of Theorem~\ref{thm:CPM}, the matrix $B_d(s)$ is invertible for all $s \in \C^+_\alpha$ and $d \in \R \setminus \{0\}$. 
\end{corollary}
\begin{proof} 
	Let $s \in \C^+_\alpha$. Then the principal square root $\sqrt{(\Cb s + \Gb)(\Lb s + \Rb)}$ exists by Corollary~\ref{cor:PSR}. Since its spectrum is contained in $\C^+$,  $\spec\left(d \sqrt{(\Cb s + \Gb)(\Lb s + \Rb)} \right) \cap \j\R = \emptyset$ for any $d \in\R \setminus \{0\}$, and so $ \sinh\left(d \sqrt{(\Cb s + \Gb)(\Lb s + \Rb)} \right)$ is invertible. 
	Moreover, $\spec(\Lb s + \Rb) \subset \C^+$ by Corollary~\ref{cor:realparts}, so $\Lb s + \Rb$ is invertible, and thus $\Zb(s)$ is invertible. \end{proof}
\smallskip
\begin{remark}[Short-circuit] \label{rem:short}
	The admittance matrix $Y(s,d)$ cannot be defined at $d=0$, because this constitutes a short-circuit.
\end{remark}
 
 \subsection{Growth bounds}
To bound the admittance matrix in \eqref{eq:admittance}, observe that $D_d(s)$ is already bounded by \eqref{eq:expbound}, as a submatrix of $\Xi(s,d)$. The challenge is then to find an upper bound on $B_d(s)^{-1}$, which via the next result, translates to a lower-bound on the determinant of $B_d(s)$.
\begin{proposition} \label{prop:mp}
	For any invertible $A \in \C^{n \times n}$,
	$$ \| A ^{-1} \| \leq \frac{\|A\|_F^{n-1}}{ |\det(A)|}.$$
\end{proposition}
\begin{proof} The bound
	$$ \sigma_n(A)  \geq \left( \frac{n-1}{\|A\|^2_F}\right) ^\frac{n-1}{2} |\det A|$$
	is proved in~\cite{gungor_erratum_2010}. Recalling $0^0 = 1$, $ k_n: = (n-1)^\frac{n-1}{2}$
	is non-decreasing over $n \geq 1$, so $k_n \geq k_1 = 1$ for all $n \geq 1$. Thus,
	$$\sigma_n(A)^ \geq \frac{ |\det A|}{\|A\|_F^{n-1}},$$ and the result follows because $\| A^{-1} \| = \sigma_n(A)^{-1}. $
	\end{proof}
Recall, now, the expression
\begin{equation}
		B_d(s)  = \Zb(s) \sinh\left(d \sqrt{(\Cb s + \Gb)(\Lb s + \Rb)} \right) \tag{\ref{eq:Bd}}
\end{equation}
in Theorem \ref{thm:CPM}. We first focus on lower-bounding the determinant of $\sinh\left(d \sqrt{(\Cb s + \Gb)(\Lb s + \Rb)} \right) $, which can be achieved by bounding the spectrum of $\sqrt{(\Cb s + \Gb)(\Lb s + \Rb)}$ away from the imaginary axis. This in turn demands bounds on the spectrum of the matrix product $(\Cb s + \Gb)(\Lb s + \Rb)$, for which there are no straightforward results (unless mutual diagonalisability of the two factors is assumed). The spectra of products of general linear operators is treated in \cite{williams_spectra_1967}, which in Theorem 1, gives
\begin{align} \sigma(A^{-1}B) &\subseteq W(B)/W(A) \label{eq:williams} \\ 
	& := \{ \tfrac{z_2}{z_1} \mid z_2 \in W(B), z_1 \in W(A) \}, \nonumber \end{align}
for complex square matrices $A$ and $B$, provided ${0 \notin W(A)}$. We therefore turn to bounds on the numerical ranges ${W((\Cb s + \Gb)^{-1})}$ and $W(\Lb s + \Rb)$.

\begin{proposition} 
\label{prop:rhpInclusion}
	For any $A \in \C^{n \times n}$,
	\begin{align*}  
		\lmin \Hr(A) & = \min \Re \big( W(A) \big)  \\
		&= \min \{ \Re(x^*Ax) \mid  x \in \C^n,\ \|x \| \geq 1 \}.
	\end{align*}
\end{proposition}
\begin{proof} For any $x \in \C^n$, 
	\begin{align*} 
		\Re( x^* A x) &=  \frac{x^*Ax + \overline{x^*Ax}}{2} \\
		& = \frac{x^*Ax + x^* A^* x}{2} = x^* \Hr(A) x. 
	\end{align*}
	Since $\Hr(A) $ is Hermitian,
	\begin{align*}
		\min \Re \big( W(A) \big)  & = \min \{ \Re(x^* Ax)  \mid  x \in \Sp^n \} \\
		& =  \min \{ x^* \Hr(A) x  \mid  x \in \Sp^n \} =\lmin \Hr(A) .
	\end{align*}
	Similarly,
	\begin{align*}
		&\min \{ \Re(x^*Ax) \mid  x \in \C^n,\ \|x \| \geq 1 \}  \\
		&= \min \{ x^* \Hr(A) x  \mid  x \in \C^n,\ \|x \| \geq 1 \}  \\
		&= \min \{ x^* \Hr(A) x  \mid  x \in \Sp^n \}  =\lmin \Hr(A).
	\end{align*}
	\end{proof}
	\smallskip

Proposition \ref{prop:rhpInclusion} encloses the numerical range within a right-half plane: $W(A) \subset \overline{\C^+_{\lmin \Hr(A)}}.$
To deal with ${W((\Cb s + \Gb)^{-1})}$ , the same must now be done for an inverse matrix.
\smallskip
\begin{lemma}
\label{lem:invRHP}
	For any invertible $A \in \C^{n \times n}$,
	$$ \{ \Re \left( z^{-1} \right) \mid z \in W(A^{-1})  \setminus \{0\} \} \subset \overline{\C^+_{\lmin \Hr(A)}}.$$
\end{lemma}
\begin{proof} 
	Suppose $z \in W(A^{-1}) \setminus \{0\}$. Then there exists $ y \in \Sp^n$ such that $z = y^* A^{-1} y$. Setting $x := A^{-1} y$, 
	$$ x^* A^* x = y^* A^{-*} A^* A^{-1} y = y^* A^{-1} y = z,$$
	so 
	$$ \frac{1}{z} = \frac{\overline{z}}{|z|^2} = \frac{\overline{x^* A^* x}}{|z|^2 } = \frac{x^* A x}{|z|^2} = w^* A w,$$ 
	where $$w := \frac{x}{|z|} = \frac{A^{-1} y}{ y^* A^{-1} y}.$$ Since $\| y \| = 1$,  the Cauchy-Schwarz implies$\|w \| \geq 1 $. Thus, ${\Re( z^{-1} )  =  \Re( w^* A w)} \geq \lmin \Hr(A)$, by Proposition~\ref{prop:rhpInclusion}.\end{proof}
	
An additional result is now required to exclude the origin from the numerical range of the inverse. 	
\begin{lemma} \label{lem:nonzero}
	For any invertible $A \in \C^{n \times n}$,
	$$ 0 \in W(A) \iff 0 \in W(A^{-1}).$$
\end{lemma}
\begin{proof} 
	Suppose $w \in W(A)$. Then $w = x^*Ax$ for some $x \in \Sp^n$. Since $A$ is invertible, $Ax \neq 0$, so set $y = \frac{Ax}{\| A x \|} \in \Sp^n$. Then
	$$ y^*A^{-1}y = \frac{ x^* A^* A^{-1} A x}{\|Ax\|^2} = \frac{x^* A^* x}{\|Ax\|^2}  = \frac{\overline{x^* A x}}{\|Ax\|^2}  = 0, $$
	by which $0 \in W(A^{-1})$.  \end{proof}
\smallskip
Before appeal to \eqref{eq:williams}, a right half-plane bound on the square root of a scalar product is required to deal with the spectrum of $\sqrt{(\Cb s + \Gb)(\Lb s + \Rb)}$.
\begin{lemma} \label{lem:geomean}
	If $w^2 = z_1 z_2$, where $z_1,z_2,w \in \overline{\C^+}$, then
	$$  \Re(w) \geq \min \{ \Re(z_1), \Re(z_2) \}.$$
\end{lemma}
\begin{proof} 
	Since $z_1,z_2 \in \overline{\C^+}$, there exist $r_1,r_2 \geq 0$ and $\theta_1, \theta_2 \in \left[ \tfrac{- \pi}{2}, \tfrac{\pi}{2} \right]$ such that $z_1 =r_1 e^{\j \theta_1} $ and  $z_2 =r_2 e^{\j \theta_2} $, so
	$ w = \sqrt{r_1r_2} e^{\j \left(\frac{ \theta_1 + \theta_2}{2}\right)} $. Clearly $\frac{ \theta_1 + \theta_2}{2} \in \left[ \tfrac{- \pi}{2}, \tfrac{\pi}{2} \right]$, by which $\cos \left( \tfrac{\theta_1 + \theta_2}{2} \right)  \geq 0 $, and therefore
	\begin{align*}
		\Re(w) = \sqrt{r_1r_2} \cos \left( \tfrac{\theta_1 + \theta_2}{2} \right) \geq  \min \{r_1,r_2\} \cos \left( \tfrac{\theta_1 + \theta_2}{2} \right).
	\end{align*}
	Since the cosine function is concave over $\left[ \tfrac{- \pi}{2}, \tfrac{\pi}{2} \right]$, 
	\begin{align*}
		\Re(w) & \geq  \min \{r_1,r_2\} \frac{\cos\theta_1 + \cos \theta_2}{2} \\
		& \geq \min \{r_1,r_2\}  \cdot \min\{  \cos\theta_1, \cos\theta_2 \},
	\end{align*}
	from which the result follows. 
	\end{proof}
\smallskip
Williams' result \eqref{eq:williams} now enables the generalisation of Lemma \ref{lem:geomean} to matrix products.
\begin{lemma} \label{lem:spectralProduct}
	If $A,B \in \C^{n \times n}$ are accretive,
	$$ \spec \sqrt{AB} \subseteq\overline{\C^+_b},$$
	where $b= \min \{ \lmin \Hr(A), \lmin \Hr(B)\}$. 
\end{lemma}
\begin{proof} 
	First recall, the existence of $\sqrt{AB}$ is guaranteed by Proposition \ref{prop:sqrts}.
	By hypothesis, $\Hr(A) \succ 0$, so ${\lmin \Hr(A) > 0}$. Proposition~\ref{prop:rhpInclusion} then implies $0 \notin W(A)$, and Lemma \ref{lem:nonzero} in turn implies $0 \notin W(A^{-1})$.
	Suppose $\lambda \in \spec \sqrt{AB}$. Then $\lambda^2 \in \spec AB$, and it follows from \cite[Theorem 1]{williams_spectra_1967}
	that $$ \exists z_1 \in W(A^{-1})\ \exists z_2 \in W(B),\ \lambda^2 = \tfrac{z_2}{z_1}.$$ 
	Observe that $\Re(z_2) \geq \lmin \Hr(B)$ by Proposition \ref{prop:rhpInclusion},  and $\Re(z_1^{-1}) \geq \lmin \Hr(A)$ by Lemma \ref{lem:invRHP}, so the result then follows from Lemma~\ref{lem:geomean}.
	\end{proof}
\smallskip

This can finally be applied to $\sqrt{(\Cb s + \Gb)(\Lb s + \Rb)}$. Below, its spectrum is bounded away from the imaginary axis when $s$ is also sufficiently far to the right.
\smallskip
\begin{lemma}
\label{lem:sqrtspec}
	Under the hypotheses of Theorem \ref{thm:CPM}, for any ${\epsilon \in \R^+ \cap [-\alpha,\infty)}$, 
	$$\bigcup_{s \in \overline{\C^+_{\alpha + \epsilon}}} \spec \sqrt{ (\Cb s + \Gb) (\Lb s + \Rb)} \subseteq \overline{\C^+_{ b \epsilon}},$$
	where $b= \min\{ \lmin( \Lb), \lmin(\Cb) \}>0$.
\end{lemma}
\begin{proof} Let $s \in \overline{\C^+_{\alpha + \epsilon}}$, where $\epsilon> 0$ is such that $\alpha + \epsilon \geq 0$. Recall from \eqref{eq:rhogamma} and \eqref{eq:alpha} that $$\rho,\gamma \leq \alpha < \alpha + \epsilon \leq \Re(s).$$
	Thus, $\Lb s + \Rb$ and $\Cb s + \Gb$ are accretive by Corollary \ref{cor:realparts}. Applying Lemma \ref{lem:scaling},
	\begin{align}
		\lmin \Hr (\Lb s + \Rb) &=  \lmin( \Lb \Re(s) + \Rb) \nonumber \\
		& \geq (\alpha + \epsilon) \lmin( \Lb )  + \lmin(\Rb)  \label{eq:Lbeta} \\
		& \geq  \epsilon \lmin(\Lb)>0, \nonumber 
	\end{align}
	because $\alpha \geq - \frac{\lmin(\Rb)}{\lmin(\Lb)}$ by \eqref{eq:alpha}.  Similarly,
	\begin{align}
		\lmin \Hr (\Cb s + \Gb) &=  \lmin( \Cb \Re(s) + \Gb) \nonumber  \\
		& \geq (\alpha + \epsilon) \lmin( \Cb )  + \lmin(\Gb)  \label{eq:Cbeta} \\
		& \geq  \epsilon \lmin(\Cb) > 0,\nonumber 
	\end{align}
	because $\alpha \geq - \frac{\lmin(\Gb)}{\lmin(\Cb)}$ by \eqref{eq:alpha}. The result then follows by Lemma~\ref{lem:spectralProduct}. 
	\end{proof}
\smallskip

As previously mentioned, the preceding right half-plane bound on the spectrum of $\sqrt{ (\Cb s + \Gb) (\Lb s + \Rb)} $ turns into a lower bound on the determinant of $\sinh\left(d \sqrt{(\Cb s + \Gb)(\Lb s + \Rb)} \right) $.

\begin{lemma} \label{lem:sinhspec}
	Under the hypotheses of Theorem \ref{thm:CPM}, let $\delta > 0$ and ${\epsilon \in \R^+ \cap [-\alpha,\infty)}.$ Then  for all $ s \in \overline{\C^+_{\alpha + \epsilon}}$ and $d \geq \delta$, 
	$$ \left|\det \sinh\left(d \sqrt{(\Cb s + \Gb)(\Lb s + \Rb)} \right) \right| \geq ( \delta  b \epsilon )^n,$$
	where $b= \min\{ \lmin( \Lb), \lmin(\Cb) \}>0$.
\end{lemma}
\begin{proof} 
	Let $s \in \overline{\C^+_{\alpha+\epsilon}}$, $d \geq \delta$, and let $\lambda_1,...,\lambda_n$ be the eigenvalues of $\sqrt{(\Cb s + \Gb)(\Lb s + \Rb)}$. Then by \cite[Theorem 1.13]{higham_functions_2008}, $\sinh (d\lambda_k)$ are the eigenvalues of $\sinh\left(d \sqrt{(\Cb s + \Gb)(\Lb s + \Rb)} \right)$, so
	\begin{align*}
		|\det \sinh\left(d \sqrt{(\Cb s + \Gb)(\Lb s + \Rb)} \right) |& = \prod_{k=1}^n |\sinh (d\lambda_k) |.
	\end{align*}
Now $\sinh$ is strictly increasing, and that $\sinh t \geq t$ for all $t \geq 0$, which is evident from its Taylor series. 
	Applying the identity $ |\sinh z|^2 = \sinh^2( \Re z) + \sin^2( \Im z)$ in \cite[(4.5.54)]{stegun_handbook_2002},
	$$| \sinh(d \lambda_k)| \geq \sinh(d|\Re\lambda_k|) \geq \sinh(\delta \epsilon_\beta) \geq \delta b \epsilon,$$ by  Lemma \ref{lem:sqrtspec}. \end{proof}
\smallskip

We finally have the desired lower bound on determinant of the right-hand factor of $B_d(s)$ in \eqref{eq:Bd}. A lower bound on the determinant of the left-hand factor $\Zb(s)$ is required next. This can be obtained by considering a generic first-order transfer matrix. 

\begin{lemma} \label{lem:rationalTFub}
	Let $A,B,P,Q \in \R^{n \times n}$, and define the transfer function $H:S \to \C^{n \times n}$ as 
	$$H(s):= (As + B)(Ps + Q)^{-1},$$
	where $S:= \{ s \in \C \mid \det(Ps+Q) \neq 0 \}$.  If $P,Q \succ 0$ then $$ \sup \{ |\det H(s) | \mid s \in \overline{\C^+} \} < \infty.$$
\end{lemma}
\begin{proof} Since ${P \succ 0}$, $Ps + Q = P(s + P^{-1}Q) $, where $P^{-1} \succ 0$. Noting that $\spec(P^{-1}Q)\subset \R^+$ by \cite[Corollary 7.6.2(a)]{horn_matrix_2012}, the set of poles of $H$ is then
	$$ \{ s \in \C \mid \det(sI + P^{-1}Q) = 0 \} = \spec(- P^{-1}Q) \subset \R^-.$$
	Thus, $H$ is a proper rational transfer matrix with poles in the open left-half plane, which implies $ H \in \Hi^{n \times n}_\infty$~\cite[Section 4.3]{zhou_robust_1996}. There exists $K > 0$ such that $\|H(s)\| \leq K$ for all $s \in \overline{\C^+}$ by Proposition \ref{prop:analcont}. It follows that $| \det H(s) | \leq K^n$, because the determinant is the product of the eigenvalues, and the spectral norm bounds the spectral radius. 
	\end{proof}
\smallskip
The previous result is an upper bound, when the denominator matrices $P,Q$ are positive definite. If the numerator matrices are as well, then we get a lower bound.
\begin{corollary} \label{cor:rationalTFlb}
	If $A,B,P,Q\succ 0$, then $$\inf \{|\det H(s)|  \mid s \in \overline{\C^+}\} > 0.$$
\end{corollary}
\begin{proof} Apply Lemma \ref{lem:rationalTFub} to $H(s)^{-1}$.
	\end{proof}
\smallskip

This result can now be applied to $\Zb(s)$, recalling its definition in \eqref{eq:Z1}. Observe, below, that $\beta = \alpha + \epsilon \geq 0$ for some $\epsilon > 0$. 
\begin{lemma} \label{lem:Zb}
	Under the hypotheses of Theorem \ref{thm:CPM}, for any ${\beta \in [0,\infty) \cap (\alpha,\infty)}$, 
	$$ \inf \{ | \det \Zb(s)| \mid s \in \overline{\C^+_\beta} \} > 0. $$ 
\end{lemma}
\begin{proof} 
	For any $s \in \C^+_\alpha$, both $\Lb s + \Rb$ and $\Cb s + \Gb$ are accretive, and therefore invertible. Thus,
	\begin{align*}
		|\det \Zb(s)| &= \frac{|\det( \Lb s + \Rb)|}{\sqrt{|\det( \Lb s + \Rb)\det( \Cb s + \Gb)|}} \\
		& = \sqrt{ \frac{|\det( \Lb s + \Rb)|}{|\det( \Cb s + \Gb)|}} \\
		&= \sqrt{ \left|\det \big( ( \Lb s + \Rb)(\Cb s + \Gb)^{-1}\big) \right|} \\
		& = \sqrt{| \det H(s) |},
	\end{align*}
	where $H(s) :=   ( \Lb (s + \beta) + \Rb)(\Cb (s + \beta) + \Gb)^{-1}.$
	Now $\Lb, \Cb \succ 0$ by Assumption \ref{ass:parameters}, and $\Lb \beta + \Rb \succ 0$ and $\Cb \beta + \Gb \succ 0$ by Lemma \ref{lem:posdef}. The result then follows from Corollary \ref{cor:rationalTFlb}. 
	\end{proof}
	Having lower-bounded the left-hand factor determinant, combining this with Lemma \ref{lem:sinhspec} yields the desired lower bound on $\det B_d(s)$. 
	\begin{corollary} \label{cor:detB}
		Under the hypotheses of Theorem \ref{thm:CPM}, for any $\delta > 0$ and $\beta \in [0,\infty) \cap (\alpha,\infty)$,
		$$ \inf \{ |\det B_d(s)| \mid s \in \overline{\C^+_\beta}, d \geq \delta\} > 0.$$
	\end{corollary}
\smallskip

A growth bound on the admittance matrix can finally be presented.

\begin{theorem}[Admittance] \label{thm:admittance}
	Under the hypotheses of Theorem~\ref{thm:CPM}, $Y: \C^+_\alpha \times \R^+ \to \C^{2n \times 2n} $ given by
	\begin{gather}
		Y(s,d) = \begin{bmatrix}
			D_d(s)B_d(s)^{-1} & -B_d(s)^{-1} \\
			-B_d(s)^{-1} & D_d(s)B_d(s)^{-1}
		\end{bmatrix} \tag{\ref{eq:admittance}}
	\end{gather}
	has the following properties: \begin{enumerate}
		\item $Y$ is continuous.
		\item For every $d \in \R^+$, the map $s \mapsto Y(s,d)$ is analytic.
		\item For any $\beta \in [0,\infty) \cap (\alpha,\infty)$ and $\delta > 0$, there exists $M > 0$ such that
		\begin{equation}
		\forall s \in \overline{\C^+_\beta}\ \forall d \geq \delta,\	\| Y(s,d) \| \leq M e^{\frac{nd}{\nu} \Re(s + \theta)}. \label{eq:Me}
		\end{equation}
	\end{enumerate} 
\end{theorem}
Recall the parameter definitions in \eqref{eq:kappa} and \eqref{eq:params}.
\begin{proof} Analyticity of both $B_d$ and $D_d$, as well as continuity with respect to $d$, follow from Theorem~\ref{thm:CPM} because they are properties of the matrix exponential. Corollary \ref{cor:Bdinv} implies $B_d(s)$ is invertible on the domain of $Y$, and matrix inversion preserves both properties, so they hold for $Y$ too. For the third property, observe the admittance matrix factors as
	$$ Y(s,d) = \left( \begin{bsmallmatrix}
		D_d(s) & 0 \\ 0 & D_d(s)
	\end{bsmallmatrix} - \begin{bsmallmatrix}
		0 & \I_n \\ \I_n & 0
	\end{bsmallmatrix}\right) \begin{bsmallmatrix}
	B_d(s)^{-1} & 0 \\ 0 & B_d(s)^{-1}
	\end{bsmallmatrix},$$
	so 
	$$ \| Y(s,d) \| \leq (\| D_d(s) \| + 1) \|B_d(s)^{-1} \|.$$
	Proposition \ref{prop:mp} and Corollary \ref{cor:detB} imply the existence of $K > 0$ such that, for all $s \in \overline{\C^+_\beta}$ and $d \geq \delta$, 
	$$ \|B_d(s)^{-1} \| \leq K \|B_d(s)\|^{n-1},$$
	recalling that $\|B_d(s)\|_F \leq \sqrt{n}\|B_d(s)\|$ by~\cite[Fact 9.8.12(ix)]{bernstein_matrix_2009}. Thus,
	\begin{align*}  
		  \| Y(s,d) \|  \leq K(\| D_d(s) \| + 1) \|B_d(s)\|^{n-1},
	\end{align*} 
	and since  $\|B_d(s)\|, \|D_d(s)\| \leq \| \Xi(s,d)\|$, 
	\begin{align*}
		\| Y(s,d) \| &\leq K \left( \| \Xi(s,d) \|^n + \| \Xi(s,d) \|^{n-1} \right).
	\end{align*}
Theorem~\ref{thm:growth} implies 
$$ \forall s \in \overline{\C^+_\beta} \ \forall d \geq \delta,\  \| \Xi(s,d) \| \leq \kappa e^{\frac{d}{\nu}(\Re (s) + \theta)}.$$
Since $\kappa \geq1$,  and $d,\nu > 0$,
$$\kappa^{n-1}e^{(n-1)\frac{d}{\nu}(\Re (s) + \theta)} \leq \kappa^n e^{n\frac{d}{\nu}(\Re (s) + \theta)},$$ and therefore $\| Y(s,d) \| \leq 2K \kappa^n  e^{n \frac{d}{\nu}(\Re (s) + \theta)}$.
	\end{proof}
	Theorem \ref{thm:admittance} suffices to establish $\mathcal{L}_2$-stability of the admittance matrix.
	\begin{corollary} \label{cor:admittanceStability}
		If $\Gb,\Rb \succ 0$, then for any $d > 0$,
		$$ \sup\{  \| Y(\j \omega,d) \| \mid \omega \in \R\} < \infty.$$
	\end{corollary}
Recalling Remark \ref{rem:lead}, the lower-bound on propagation speed $\frac{\nu}{n}$ implied by Theorem \ref{thm:admittance} is more conservative than the lower-bound $\nu$ implied by Theorem~\ref{thm:growth}. 
\section{The impedance matrix} \label{sec:impedance}
Given the transmission line model in Section \ref{sec:DPmodel}, consider a \emph{dual line} with inductance $\Lb':= \Cb$, capacitance $\Cb':= \Lb$, resistance $\Rb':= \Gb$ and conductance $\Gb':= \Rb$. From Theorem~\ref{thm:CPM}, it is clear that the ABCD matrices of the dual are related to those of the primal via
\begin{align*}
	&A'_d(s) = D_d(s), &B'_d(s) = C_d(s), \\ &C'_d(s) = B_d(s), & D'_d(s) = A_d(s).
\end{align*}
The impedance matrix of the primal is given by
\begin{equation*}
	Z(s,d) = \begin{bmatrix}
		A_d(s) C_d(s)^{-1} & A_d(s) C_d(s)^{-1} D_d(s) - B_d(s) \\ C_d(s)^{-1} & C_d(s)^{-1} D_d(s)
	\end{bmatrix},
\end{equation*}
in terms of its ABCD parameters~\cite[(9)]{reveyrand_multiport_2018}.
Comparing this with \eqref{eq:primalAdmittance} reveals that $$Z(s,d) = \begin{bsmallmatrix}
	0 & -\I_n \\ \I_n & 0
\end{bsmallmatrix}Y'(s,d)\begin{bsmallmatrix}
0 & -\I_n \\ \I_n & 0
\end{bsmallmatrix},$$
by which $ \|Z(s,d) \| = \|Y'(s,d) \|$ for all $s$ and $d$.
\begin{remark}[Duality] \label{rem:dual}
	The impedance matrix of a transmission line is unitarily similar to the admittance matrix of its dual.
\end{remark}
Impedance matrices thus share all the properties established in Theorem \ref{thm:admittance} for admittance matrices. This is stated formally below.
 \begin{theorem}[Telegrapher's impedance matrix] \label{thm:impedance}
 	Under the hypotheses of Theorem \ref{thm:CPM}, the impedance matrix ${Z: \C^+_\alpha \times \R^+ \to \C^{2n \times 2n}}$ given by
 	$Z(s,d):= Y(s,d)^{-1}$, has the following properties: \begin{enumerate}
 		\item $Z$ is continuous.
 		\item For every $d \in \R^+$, the map $s \mapsto Z(s,d)$ is analytic.
 		\item For any $\beta \in [0,\infty) \cap (\alpha,\infty)$ and $\delta > 0$, there exists $M > 0$ such that
 		\begin{equation*}
 			\forall s \in \overline{\C^+_\beta}\ \forall d \geq \delta,\	\| Z(s,d) \| \leq M e^{\frac{nd}{\nu} \Re(s + \theta)}.
 		\end{equation*}
\end{enumerate}
\end{theorem}
\begin{proof}
In view of Remark \ref{rem:dual}, it only remains to show that the dual $\kappa', \theta', \nu'$ are identical in value to the primal $\kappa, \theta, \nu$. 
	Since $$ \begin{bmatrix}
		0 & \Lb \\ \Cb & 0
	\end{bmatrix} = \begin{bmatrix}
		0 & \I_n \\ \I_n & 0
	\end{bmatrix}\begin{bmatrix}
		0 & \Cb \\ \Lb & 0
	\end{bmatrix} \begin{bmatrix}
		0 & \I_n \\ \I_n & 0
	\end{bmatrix},$$
	it follows that $e^{ \left( \j \omega \begin{bsmallmatrix}
			0 & \Lb \\ \Cb & 0
		\end{bsmallmatrix} \right)}$ and $e^{ \left( \j \omega \begin{bsmallmatrix}
			0 & \Lb' \\ \Cb' & 0
		\end{bsmallmatrix}\right)}$ are unitarily similar, and so share the same spectral norm. Equation \eqref{eq:kappa} then implies $\kappa'=\kappa$, by which $(\theta',\nu') = (\theta,\nu)$ from \eqref{eq:params}. 
\end{proof}
	\begin{corollary} \label{cor:impedanceStability}
	If $\Gb,\Rb \succ 0$, then for any $d > 0$,
	$$ \sup\{  \| Z(\j \omega,d) \| \mid \omega \in \R\} < \infty.$$
\end{corollary}
\section{Conclusion}
\label{concl}
The chain, ABCD, admittance and impedance matrices of a \textcolor{black}{multiconductor} transmission line have been derived, as transfer functions parametrised by the length of the line. This relies on the linearity and time-invariance of the underlying dynamics~\cite[Chapter 3.4.1]{dullerud_course_2000}. Mathematically, each transfer function corresponds to a different linear operator, with different pairs of voltages and currents designated as inputs and outputs. None of the transfer functions are rational, implying they all describe infinite-dimensional dynamics. This is consistent with their derivation from the Telegrapher's equation, a linear time-invariant PDE. As a result, this line model exhibits a finite speed of information propagation, a feature that cannot be reproduced by any finite-dimensional model. 

\textcolor{black}{
Frequency-domain bounds for each of the transfer matrices have also been derived. The bounds have implications for network stability, causality, and transmission delays. The chain and ABCD matrices are bounded over the imaginary axis, which establishes their $\mathcal{L}_2$-stability. This means the energy of the output signals is bounded linearly by the energy of the input signals.
The impedance and admittance matrices are, likewise, shown to be $\mathcal{L}_2$-stable when the line conductance and resistance matrices are positive-definite. } 

An $\mathcal{L}_2$-stable LTI system is causal if and only if its transfer function lies in $\mathcal{H}_\infty$~\cite[Chapter 3.4.3]{dullerud_course_2000}, which requires boundedness over the complex right half-plane. \textcolor{black}{However, the bounds derived here permit all four transfer matrices to grow exponentially as $\Re(s) \to \infty$. The chain, ABCD, admittance and impedance matrices of a distributed parameter transmission line are, in fact, not causal. This is due to the bidirectional flow of information between inputs and outputs. The finite speed of information propagation is lower-bounded by the rate of exponential growth in the frequency-domain.}

\textcolor{black}{
The physical properties mentioned above could be expected of a transmission line on purely physical grounds. Our contribution is to demonstrate that they are system-theoretic consequences of using the Telegrapher's equation to model the line.
We envisage fault localisation as a potential domain of application for the results. The transfer matrix expressions found here can be used to build models for transmission networks. Removal of the diagonalisability assumption is important, because it cannot be guaranteed under fault conditions. Our bounds over the complex plane ensure that Fourier transforms can be invoked to predict the response of the grid to faults. Propagation delays between the fault and the sensors can also be accounted for, by appeal to the bounds. The details of such an approach are developed in ~\cite{selvaratnam_fault_2025}.}

\section*{Acknowledgements}
\textcolor{black}{
The authors thank J. A. Brand\~{a}o Faria for his thoughtful interaction with an initial draft of this work. His comments helped to clarify and enhance our contributions.}
\printbibliography

\end{document}